\documentstyle[12pt]{article}
\begin{document}
\baselineskip 0.6cm

\def\simgt{\mathrel{\lower2.5pt\vbox{\lineskip=0pt\baselineskip=0pt
           \hbox{$>$}\hbox{$\sim$}}}}
\def\simlt{\mathrel{\lower2.5pt\vbox{\lineskip=0pt\baselineskip=0pt
           \hbox{$<$}\hbox{$\sim$}}}}
\def\lrD{\stackrel{\leftrightarrow}{\partial}}

\begin{titlepage}

\begin{flushright}
UCB-PTH-05/27 \\
LBNL-58857
\end{flushright}

\vskip 2.0cm

\begin{center}

{\Large \bf 
Minimally Fine-Tuned Supersymmetric Standard Models \\
with Intermediate-Scale Supersymmetry Breaking
}

\vskip 1.0cm

{\large
Yasunori Nomura, David Poland, and Brock Tweedie
}

\vskip 0.4cm

{\it Department of Physics, University of California,
           Berkeley, CA 94720} \\
{\it Theoretical Physics Group, Lawrence Berkeley National Laboratory,
           Berkeley, CA 94720} \\

\vskip 1.2cm

\abstract{
We construct realistic supersymmetric theories in which the correct 
scale for electroweak symmetry breaking is obtained without significant 
fine-tuning.  We consider two classes of models.  In one class 
supersymmetry breaking is transmitted to the supersymmetric standard 
model sector through Dirac gaugino mass terms generated by a $D$-term 
vacuum expectation value of a $U(1)$ gauge field.  In the other class 
the supersymmetry breaking sector is separated from the supersymmetric 
standard model sector in an extra dimension, and the transmission of 
supersymmetry breaking occurs through gauge mediation.  In both these 
theories the Higgs sector contains two Higgs doublets and a singlet, 
but unlike the case for the next-to-minimal supersymmetric standard 
model the singlet field is not responsible for generating the 
supersymmetric or supersymmetry breaking mass for the Higgs doublets. 
These masses, as well as the mass for the singlet, are generated through 
gravitational-strength interactions.  The scale at which the squark and 
slepton masses are generated is of order $(1\!\sim\!100)~{\rm TeV}$, 
and the generated masses do not respect the unified mass relations. 
We find that electroweak symmetry breaking in these theories is caused 
by an interplay between the top-stop radiative correction and the 
holomorphic supersymmetry breaking mass for the Higgs doublets and 
that the fine-tuning can be reduced to the level of $20\%$.  The 
theories have rich phenomenology, including a variety of possibilities 
for the lightest supersymmetric particle.}

\end{center}
\end{titlepage}

\section{Introduction}
\label{sec:intro}

Weak scale supersymmetry is an attractive candidate for physics 
beyond the standard model.  It not only stabilizes the Higgs potential 
against potentially large radiative corrections, but also leads to an 
elegant picture of gauge coupling unification at a scale $M_X \simeq 
10^{16}~{\rm GeV}$~\cite{Dimopoulos:1981zb}.  Non-discovery of the 
superparticles or the Higgs boson at LEP~II, however, puts strong 
constraints on how supersymmetry can be realized at the weak scale. 
In the minimal supersymmetric standard model (MSSM), evading the lower 
bound on the physical Higgs boson mass typically requires a large top squark 
mass.  This in turn gives a large radiative correction to the Higgs boson 
mass-squared parameter, leading to fine-tuning in electroweak symmetry 
breaking.  In fact, the problem of fine-tuning is somewhat generic in 
theories with weak scale supersymmetry, and is called the supersymmetric 
fine-tuning problem. 

Recently, a general framework has been discussed for supersymmetric 
theories that avoid fine-tuning while preserving the successful features 
of supersymmetry~\cite{Chacko:2005ra,Nomura:2005qg}.  The key point is to 
lower both the top squark masses and the mediation scale of supersymmetry 
breaking, by violating simple unified mass relations.%
\footnote{The unified mass relations need not be violated if 
the top quark and Higgs boson are both rather heavy, $m_t \simeq 
(180\!\sim\!182)~{\rm GeV}$ and $M_{\rm Higgs} \simeq 
(200\!\sim\!250)~{\rm GeV}$~\cite{Nomura:2005qg}.  We do not 
consider this scenario in this paper.}
This makes radiative corrections to the Higgs mass-squared parameter 
small, and thus reduces fine-tuning.  A simple way to accommodate 
such light top squarks is to introduce an additional contribution 
to the Higgs quartic couplings other than that from the $SU(2)_L 
\times U(1)_Y$ $D$-terms of the MSSM (for theories giving such 
a contribution, see e.g.~[\ref{Ellis:1988er:X}~--~\ref{Babu:2004xg:X}]).%
\footnote{A solution to the supersymmetric fine-tuning problem that 
does not require an extension of the Higgs sector at the weak scale has 
recently been presented in~\cite{Kitano:2005wc}, where the supersymmetry 
breaking mass for the up-type Higgs boson is suppressed by a cancellation 
between two different contributions~\cite{Choi:2005uz}, and a large stop 
mixing parameter and a small holomorphic supersymmetry breaking Higgs 
mass ensure successful electroweak symmetry breaking with relatively 
small top squark masses.}
In Refs.~\cite{Chacko:2005ra,Nomura:2005qg} these were achieved 
by adopting the mechanism of~\cite{Nomura:2004is}: the dynamical 
supersymmetry breaking sector has a global $SU(5)$ symmetry, of which 
the $SU(3) \times SU(2) \times U(1)$ subgroup is explicitly gauged 
and identified as the standard model gauge group, but this $SU(5)$ 
is spontaneously broken to the gauged subgroup at the dynamical 
scale of $\Lambda \approx (10\!\sim\!100)~{\rm TeV}$.  This structure 
allows us to accommodate the successful prediction associated with 
gauge coupling unification at the leading-log level, while the unwanted 
unified mass relations among the squark and slepton masses are avoided. 
The mediation scale of supersymmetry breaking is very low and of order 
$\Lambda$, which is the scale of dynamical supersymmetry breaking itself.
The Higgs sector superpotential is generated from interactions between 
the Higgs and dynamical supersymmetry breaking sectors through marginal 
operators.

In this paper we construct classes of explicit supersymmetric standard 
models in which the fundamental scale of supersymmetry breaking is 
an intermediate scale, $M_I = \sqrt{m_{\rm weak} M_{\rm Pl}}$, and yet 
the supersymmetric fine-tuning problem is ameliorated.  An advantage 
of intermediate-scale supersymmetry breaking is that the Higgs sector 
superpotential is obtained relatively easily through nonrenormalizable 
interactions suppressed by the Planck scale~\cite{Giudice:1988yz}.  To 
implement this mechanism without introducing the supersymmetric flavor 
problem, we consider two classes of theories.  In one class supersymmetry 
breaking is transmitted to the supersymmetric standard model sector 
through a $D$-term vacuum expectation value (VEV) of a $U(1)$ gauge 
field~\cite{Fox:2002bu}.  In the other class the supersymmetry breaking 
sector is separated from the supersymmetric standard model sector in an 
extra dimension, and the transmission of supersymmetry breaking occurs 
through gauge mediation~\cite{Nomura:2000uw}.  In both these theories 
the Higgs sector contains two Higgs doublets and a singlet.  Our Higgs 
sector superpotential, however, differs from that of the next-to-minimal 
supersymmetric standard model: it contains weak-scale mass parameters 
which are naturally generated through gravitational-strength interactions. 
The scale at which the squark and slepton masses are generated is of order 
$(1\!\sim\!100)~{\rm TeV}$, and the generated masses do not respect the 
unified mass relations.  These features nicely meet the general criteria 
discussed above.  (Note that the relevant scale for the fine-tuning 
argument is the scale at which the squark and slepton masses are 
generated, and not the one at which the gaugino masses are generated.) 

Electroweak symmetry breaking in our theories occurs because of an interplay 
between the top-stop radiative correction and the holomorphic supersymmetry 
breaking mass squared for the Higgs doublets, the $\mu B$ term.  The 
dynamics of the singlet field $S$ almost decouples from the electroweak 
symmetry breaking physics due to a relatively large supersymmetric mass 
for $S$.  These theories, therefore, naturally realize the scenario~II 
of $\mu B$-driven electroweak symmetry breaking discussed in 
Ref.~\cite{Nomura:2005rk}.  This has an advantage, compared with 
theories based on the next-to-minimal supersymmetric standard model 
(NMSSM), that there is no strict requirement on the potential that the 
correct supersymmetric Higgs mass term ($\mu$ term) should be reproduced 
by the VEV of the singlet field, thus opening up a larger region of 
parameter space that correctly breaks the electroweak symmetry.  The 
$\mu$ and $\mu B$ terms of order the weak scale are naturally generated 
in our theories through gravitational-strength interactions.  (For 
attempts of reducing fine-tuning in the context of the NMSSM, see 
e.g.~\cite{Bastero-Gil:2000bw}.)  In the present scheme, the amount 
of fine-tuning is essentially determined by the ratio of the lightest 
neutral Higgs-boson and the charged Higgs-boson squared masses.  We 
find that the fine-tuning in these theories can be reduced to the 
level of $20\%$. 

The organization of the paper is as follows.  In section~\ref{sec:model-1} 
we discuss the first class of theories.  We study electroweak symmetry 
breaking and the superparticle spectrum, identifying some characteristic 
features of the theories.  In section~\ref{sec:model-2} we discuss the 
second class of theories and perform a similar analysis of electroweak 
symmetry breaking and the superparticle spectrum.  Conclusions are 
given in section~\ref{sec:concl}.

\section{Models with {\boldmath $D$}-type Supersymmetry Breaking}
\label{sec:model-1}

In this section we present the first class of models, in which 
supersymmetry breaking is transmitted to the supersymmetric standard 
model sector through a $D$-term VEV of a $U(1)$ gauge field.  We find 
that the fine-tuning is reduced to the level of $20\%$.

\subsection{Supersymmetry breaking from a {\boldmath $D$}-term VEV}
\label{subsec:1-susy-br}

The supersymmetric standard model sector of our theories contains, 
as usual, the $SU(3)_C \times SU(2)_L \times U(1)_Y$ (321) gauge 
multiplet, $V_i$ ($i=1,2,3$ for $U(1)_Y$, $SU(2)_L$ and $SU(3)_C$), 
and three generations of matter fields, $Q$, $U$, $D$, $L$ and $E$. 
We also introduce a gauge singlet chiral superfield $S$ as well 
as two Higgs doublets $H_u$ and $H_d$, with the standard Yukawa 
couplings in the superpotential $W = y_u Q U H_u + y_d Q D H_d 
+ y_e L E H_d$. 

Following Ref.~\cite{Fox:2002bu}, we consider that supersymmetry 
breaking is transmitted to the supersymmetric standard model sector 
through a $D$-term VEV of a $U(1)$ gauge interaction, $U(1)'$:
\begin{equation}
  \langle V' \rangle = \frac{1}{2} \theta^2 \bar{\theta}^2 D',
\label{eq:D-U1'}
\end{equation}
where $V'$ is the vector superfield for $U(1)'$.  Introducing 
chiral superfields that transform as adjoints under 321, $A_1({\bf 1}, 
{\bf 1})_0$, $A_2({\bf 1}, {\bf 3})_0$ and $A_3({\bf 8}, {\bf 1})_0$, 
supersymmetry breaking in Eq.~(\ref{eq:D-U1'}) is transmitted to the 
supersymmetric standard model sector through the following operators:
\begin{equation}
  {\cal L} = \sum_{i=1,2,3} \int\!d^2\theta\, \frac{\zeta_i}{M_*} 
     {\cal W}_i^\alpha {\cal W}'_\alpha A_i + {\rm h.c.},
\label{eq:gaugino-mass}
\end{equation}
where $\zeta_i$ are coefficients of $O(1)$, $M_*$ is a mass parameter 
of order the Planck scale, and ${\cal W}_i^\alpha$ and ${\cal W}'^\alpha$ 
are the field-strength superfields for 321 and $U(1)'$, respectively.%
\footnote{We assume that $D'$ is much larger than the largest $F$-type 
VEV, $F$, in the theory, i.e. $D' \gg F$, so that the contributions 
to the supersymmetry breaking masses from $F$ are negligible.  For a 
discussion on how to obtain $D' \gg F$, see e.g.~\cite{Gregoire:2005jr}. 
Alternatively, one can separate the field giving the largest $F$ 
from the supersymmetric standard model sector in an extra dimension; see 
e.g.~\cite{Carpenter:2005tz}.  (Mediation of supersymmetry breaking by 
Eq.~(\ref{eq:gaugino-mass}) was also considered in~\cite{Chacko:2004mi,%
Nomura:2005qg} in a slightly different context, in which the $A_i$ 
fields arise as composites.)}
These operators generate Dirac masses for the gauginos of order $D'/M_*$, 
which in turn generate flavor-universal squark and slepton squared 
masses of order $(1/16\pi^2)(D'/M_*)^2$ at one loop.  With $M_*$ of 
order the Planck scale, $D' \sim (10^{10}{\rm -}10^{11}~{\rm GeV})^2$. 
An important property of this transmission is that the squark and 
slepton masses are generated at the scale of Dirac gaugino masses 
$\approx D'/M_* \sim (1{\rm -}10)~{\rm TeV}$, although the gaugino 
masses are present at the scale $M_*$.  This reduces the logarithm 
associated with the top-Yukawa induced radiative correction to the 
Higgs soft supersymmetry breaking mass, and thus helps the reduction 
of fine-tuning.

We take the coefficients $\zeta_i$ in Eq.~(\ref{eq:gaugino-mass}) 
to be free parameters.  In particular, we do not impose any unified 
relations on the three coefficients $\zeta_1$, $\zeta_2$ and $\zeta_3$. 
This is necessary to break unwanted unified mass relations for the 
squarks and sleptons, such as $m_{\tilde{t}}^2/m_{\tilde{e}}^2 \approx 
(4 g_3^4/3)/(3 g_1^4/5)$, and to reduce fine-tuning.  (For $\zeta_1 
= \zeta_2 = \zeta_3$ in the basis where the gauge kinetic terms 
are given by ${\cal L} = \sum_i \int\!d^2\theta\, (1/4 g_i^2) 
{\cal W}_i^\alpha {\cal W}_{i\alpha} + {\rm h.c.}$, which is 
expected to be the case in naive unified theories, the squarks and 
sleptons obey unwanted unified mass relations $m_{\tilde{f}}^2 \propto 
\sum_i g_i^4 C_i^{\tilde{f}}$, where $\tilde{f} = \tilde{q}, \tilde{u}, 
\tilde{d}, \tilde{l}, \tilde{e}$ and $C_i^{\tilde{f}}$ are the group 
theoretical factors.)

The introduction of the $A_i$ fields destroys the successful supersymmetric 
prediction for gauge coupling unification.  Unification of the couplings 
can be recovered by the introduction of arbitrary vector-like matter 
fields, but at the price of losing the predictivity for the low-energy 
gauge couplings.  A possibility of recovering the prediction is to use 
the trinification idea, which has been discussed in~\cite{Fox:2002bu}. 
(The $SU(5)$ case leads to a Landau pole for the gauge couplings much 
below the unification scale at two loops.)  This issue does not occur 
in the model presented in section~\ref{sec:model-2}, and we do not 
discuss it further in the context of the present model.

\subsection{Masses for the {\boldmath $A$} fields}
\label{subsec:1-A-masses}

For $D' \sim (10^{10}{\rm -}10^{11}~{\rm GeV})^2$, the gravitino mass 
is roughly of the order of the weak scale: $m_{3/2} \sim D'/M_{\rm Pl}$, 
where $M_{\rm Pl}$ is the reduced Planck scale.  The precise value 
of $m_{3/2}$ depends on various unknown parameters, for example on 
$M_*/M_{\rm Pl}$, so here we take $m_{3/2}$ to be a free parameter 
of order $m_{3/2} \sim (100~{\rm GeV}{\rm -}1~{\rm TeV})$.  With these 
values of $m_{3/2}$, supersymmetric masses of order the weak scale 
can be naturally generated through the K\"ahler potential.  For 
example, if the K\"ahler potential contains the term $K = \sum_i 
\lambda_{A_i} A_i^2/2 + {\rm h.c.}$ ($i=1,2,3$), where $\lambda_{A_i}$ 
are dimensionless coefficients, the supergravity Lagrangian produces 
the effective superpotential term 
\begin{equation}
  W_{\rm eff} = \frac{1}{2} \sum_{i=1,2,3} m_{A,i} A_i^2,
\label{eq:SUSY-mass-Ai}
\end{equation}
where $m_{A,i} = \lambda_{A_i} m_{3/2}$~\cite{Giudice:1988yz}. 
This can be understood easily in the compensator formalism (see 
e.g.~\cite{Randall:1998uk}), in which the above K\"ahler potential 
term can be written as 
\begin{equation}
  {\cal L} = \int\!d^4\theta\, \frac{\phi^\dagger}{\phi} 
    \sum_{i=1,2,3} \frac{\lambda_{A_i}}{2} A_i^2 + {\rm h.c.},
\label{eq:Kahler-Ai}
\end{equation}
in the normalization where $A_i$ are canonically normalized.  Here, 
$\phi$ is the compensator field, which takes the value $\phi = 
1 + \theta^2 m_{3/2}$.  We then find that Eq.~(\ref{eq:Kahler-Ai}) 
gives the supersymmetric mass term of Eq.~(\ref{eq:SUSY-mass-Ai}) as 
well as the soft supersymmetry breaking term ${\cal L}_{\rm soft} 
= - \sum_i m_{A,i} m_{3/2} \, a_i^2/2 + {\rm h.c.}$, where $a_i$ are 
the lowest components of $A_i$.  Note that, assuming real couplings, 
we still have a freedom of choosing the signs of $m_{A,i}$ (in the 
phase convention that $\zeta_i$ and $m_{3/2}$ are real and positive). 
Soft supersymmetry breaking terms could also receive contributions 
from the operators
\begin{equation}
  {\cal L} = \sum_{i=1,2,3} \int\!d^2\theta\, \frac{\eta_i}{2 M_*^2} 
     {\cal W}'^\alpha {\cal W}'_\alpha\, A_i^2 + {\rm h.c.}.
\label{eq:muB-Ai}
\end{equation}
Together with the contributions from the K\"ahler potential of 
Eq.~(\ref{eq:Kahler-Ai}), we obtain 
\begin{eqnarray}
  {\cal L}_{\rm soft} &=& - \frac{1}{2} \sum_{i=1,2,3} 
    \biggl( m_{A,i} m_{3/2} - \frac{\eta_i D'^2}{M_*^2} \biggr)\, 
    a_i^2 + {\rm h.c.}
\nonumber\\
  &\equiv& - \frac{1}{2} \sum_{i=1,2,3} b_{A,i} a_i^2 + {\rm h.c.}.
\label{eq:soft-mass-Ai}
\end{eqnarray}
Here, as in the case of $\zeta_i$, we do not impose any unified 
relations on $\lambda_{A_i}$ or $\eta_i$.  The conditions for quadratic 
stability of the $a_i$-field origin are given by
\begin{equation}
  |m_{A,i}|^4 + 4g_i^2 |m_{D,i}|^2 |m_{A,i}|^2 - |b_{A,i}|^2
    + 2 g_i^2 (b_{A,i} m_{D,i}^{*2} + b_{A,i}^* m_{D,i}^2) > 0,
\label{eq:Ai=0}
\end{equation}
where $m_{D,i} \equiv -i \zeta_i D'/\sqrt{2} M_*$.  We assume that 
Eq.~(\ref{eq:Ai=0}) are satisfied for all $i=1,2,3$.

The introduction of a gauge singlet $A_1$ has a potential danger 
of destabilizing the gauge hierarchy.  Specifically, the operator 
${\cal L} \approx \int\!d^4\theta\, M_* \phi^\dagger A_1$, together 
with Eq.~(\ref{eq:SUSY-mass-Ai}), could lead to a large VEV for $A_1$, 
and thus to a large Fayet-Iliopoulos $D$ term for $U(1)_Y$ through 
Eq.~(\ref{eq:gaugino-mass}).%
\footnote{The direct kinetic mixing between $U(1)'$ and $U(1)_Y$, 
${\cal L} = \int\!d^2\theta\, {\cal W}_1^\alpha {\cal W}'_\alpha$, 
is assumed to be absent throughout.}
Even if absent at tree level, this operator would be generated at 
radiative level under the presence of general nonrenormalizable 
operators.  We avoid this by imposing a symmetry
\begin{equation}
  V' \leftrightarrow -V', \qquad A_i \leftrightarrow -A_i,
\label{eq:V'-parity}
\end{equation}
on the interactions of the observable sector.  This symmetry is broken 
by the $D'$ and physics generating it.  However, if the breaking appears 
sufficiently soft in the observable sector, the dangerous operator linear 
in $A_1$ is sufficiently suppressed.  Such a setup can naturally arise, 
for example, by generating $D'$ on the infrared brane in warped space 
(with the infrared-brane scale set to $\approx D'$) and transmitting 
it to the observable sector on the ultraviolet brane through a bulk 
$U(1)'$.  In the following, we assume that the operator linear in $A_1$ 
is sufficiently suppressed.

\subsection{The Higgs sector}
\label{subsec:1-Higgs}

The Higgs sector of our theory consists of three chiral superfields 
$S({\bf 1}, {\bf 1})_0$, $H_u({\bf 1}, {\bf 2})_{1/2}$ and $H_d({\bf 1}, 
{\bf 2})_{-1/2}$.  There are some variations on possible interactions in 
the Higgs sector.  Here, to demonstrate our point, we adopt a particular 
setup that uses a discrete $Z_{4,R}$ symmetry to constrain the form 
of these interactions. 

We consider a discrete $R$ symmetry, $Z_{4,R}$, under which fields 
transform as in Table~\ref{table:Z4R}.
\begin{table}
\begin{center}
\begin{tabular}{|c|cc|ccccc|ccc|c|}  \hline 
 & $V_i$ & $A_i$ & $Q$ & $U$ & $D$ & $L$ & $E$ & $H_u$ & $H_d$ & 
   $S$ & $V'$ \\ \hline
 $Z_{4,R}$ & $0$ & $0$ & $1$ & $1$ & $1$ & $1$ & $1$ & $0$ & $0$ & 
   $2$ & $0$
\\ \hline
\end{tabular}
\end{center}
\caption{$Z_{4,R}$ charges for the fields.}
\label{table:Z4R}
\end{table}
This charge assignment allows all the interactions discussed so far, 
including the Yukawa couplings and Eq.~(\ref{eq:gaugino-mass}).%
\footnote{This $Z_{4,R}$ symmetry forbids dangerous dimension four 
and five proton decay operators, as well as a large tree-level 
supersymmetric mass for the Higgs doublets.  It is broken by the 
VEV of the compensator field (a constant term in the superpotential 
needed to cancel the cosmological constant) to the $Z_{2,R}$ subgroup, 
which is nothing but the standard $R$ parity.}
In the absence of supersymmetry breaking, the Higgs sector superpotential 
consistent with $Z_{4,R}$ is $W_0 = \lambda S H_u H_d + (\kappa/3) S^3$. 
(We assume that the possible term linear in $S$ is absent.)  In addition, 
we have terms arising from the K\"ahler potential $K = \lambda_H H_u H_d 
+ (\lambda_S/2) S^2 + {\rm h.c.}$.  Adding these together, the 
superpotential of our Higgs sector is given by
\begin{equation}
  W_H = \lambda S H_u H_d + \mu H_u H_d 
    + \frac{M_S}{2} S^2 + \frac{\kappa}{3} S^3,
\label{eq:W_H}
\end{equation}
where $\mu = \lambda_H m_{3/2}$ and $M_S = \lambda_S m_{3/2}$ are 
mass parameters of order the weak scale.%
\footnote{The superpotential of Eq.~(\ref{eq:W_H}) can also be 
written in the form of $W_H = \lambda S H_u H_d + f(S)$, where $f(S)$ 
is a general cubic function of $S$, by shifting the $S$ field as 
$S \rightarrow S - \mu/\lambda$.}
Soft supersymmetry breaking parameters arise from the K\"ahler potential 
terms as well as from the operators ${\cal L} = \int\!d^2\theta\, 
(\eta_H H_u H_d + \eta_S S^2/2) {\cal W}'^\alpha {\cal W}'_\alpha/M_*^2 
+ {\rm h.c.}$, giving
\begin{eqnarray}
  {\cal L}_{H,{\rm soft}} &=& 
    - \biggl( \mu\, m_{3/2} - \frac{\eta_H D'^2}{M_*^2} \biggr)\, 
    H_u H_d - \frac{1}{2} \biggl( M_S m_{3/2} 
    - \frac{\eta_S D'^2}{M_*^2} \biggr)\, S^2 + {\rm h.c.}
\nonumber\\
  &\equiv& - b_H H_u H_d - \frac{b_S}{2} S^2 + {\rm h.c.},
\label{eq:soft-mass-H-S}
\end{eqnarray}
where we have used the same symbol for a chiral superfield and its 
scalar component for $H_u$, $H_d$ and $S$.  The Higgs doublets also 
obtain non-holomorphic supersymmetry breaking masses at one loop 
through 321 gauge interactions.

\subsection{Parameters at the weak scale}
\label{subsec:1-para}

Contributions to the gaugino masses arise from the operators in 
Eqs.~(\ref{eq:gaugino-mass},~\ref{eq:SUSY-mass-Ai}).  The masses of 
adjoint scalars also come from Eq.~(\ref{eq:soft-mass-Ai}).  Defining 
component fields as $V = - \theta^\alpha \sigma^\mu_{\alpha\dot{\alpha}} 
\bar{\theta}^{\dot{\alpha}} A_\mu - i \bar{\theta}^2 \theta^\alpha 
\lambda_\alpha + i \theta^2 \bar{\theta}_{\dot{\alpha}} 
\lambda^{\dagger\dot{\alpha}} + (1/2)\theta^2 \bar{\theta}^2 D$ 
and $A = a + \sqrt{2}\theta^\alpha \psi_\alpha + \theta^2 F$, 
these operators give
\begin{eqnarray}
  {\cal L} &=& - m_D \lambda^\alpha \psi_\alpha 
    - m_D^* \lambda^\dagger_{\dot{\alpha}} \psi^{\dagger \dot{\alpha}}
    + \sqrt{2} i\, m_D D a - \sqrt{2} i\, m_D^* D a^\dagger
\nonumber\\
  && {} - \frac{1}{2} m_A \psi^\alpha \psi_\alpha - \frac{1}{2} 
      m_A^* \psi^\dagger_{\dot{\alpha}} \psi^{\dagger \dot{\alpha}}
    + m_A a F + m_A^* a^\dagger F^\dagger 
\nonumber\\
  && {} - \frac{1}{2} b_A a^2 - \frac{1}{2} b_A^* a^{\dagger 2}
    - m_a^2 a^\dagger a,
\label{eq:operators}
\end{eqnarray}
for each gauge group factor $SU(3)_C$, $SU(2)_L$ and $U(1)_Y$ (we have 
suppressed the index $i=1,2,3$).  Here, we have added non-holomorphic 
supersymmetry breaking masses for $a$'s (the last term), although they 
are small in the present theory.  The mass parameters $m_D \equiv 
-(i \zeta_i/\sqrt{2}) \langle D' \rangle/M_*$, $m_A \equiv m_{A,i}$ and 
$b_A \equiv b_{A,i}$ are of order the weak scale or somewhat (an order 
of magnitude) larger.  The normalizations for $A_\mu$, $\lambda$ and $D$ 
are such that the inverse squares of the 4D gauge couplings, $1/g_i^2$, 
appear in front of the kinetic terms. 

We assume that the parameters in Eq.~(\ref{eq:operators}) are real. 
There are two gauginos, mixtures of $\lambda$ and $\psi$, for each 
gauge group factor, and their masses are given by diagonalizing 
Eq.~(\ref{eq:operators}) as
\begin{equation}
  m_{\lambda}^2 = \frac{1}{2} \left\{ 2 g^2 m_D^2 + m_A^2 
    \pm \sqrt{4 g^2 m_D^2 m_A^2 + m_A^4} \right\},
\label{eq:gaugino-explicit}
\end{equation}
where we have suppressed the index $i=1,2,3$ for $m_{\lambda}$, $g$, 
$m_D$ and $m_A$.  The squark and slepton masses arise from finite 
one-loop diagrams as
\begin{equation}
  m_{\tilde{f}}^2 = \sum_{i=1,2,3} \frac{g_i^4 C_i^{\tilde{f}}}{4 \pi^2}\, 
    \hat{M}_i^2,
\label{eq:sfermion-explicit}
\end{equation}
where $(C_1^{\tilde{f}}, C_2^{\tilde{f}}, C_3^{\tilde{f}}) = 
(1/60,3/4,4/3)$, $(4/15,0,4/3)$, $(1/15,0,4/3)$, $(3/20,3/4,0)$ and 
$(3/5,0,0)$ for $\tilde{f} = \tilde{q}, \tilde{u}, \tilde{d}, \tilde{l}$ 
and $\tilde{e}$, respectively, and $\hat{M}_i^2$ are given by
\begin{equation}
  \hat{M}_i^2 = m_D^2 
    \Biggl\{ \ln \Biggl( \frac{4 g^2 m_D^2 + m_A^2 - b_A + m_a^2}
    {g^2 m_D^2} \Biggr) - \frac{m_A}{\sqrt{4 g^2 m_D^2+m_A^2}} 
    \ln \Biggl( \frac{\sqrt{4 g^2 m_D^2 + m_A^2} + m_A}
    {\sqrt{4 g^2 m_D^2 + m_A^2} - m_A} \Biggr) \Biggr\},
\label{eq:hat-M2}
\end{equation}
which are positive in the entire parameter region~\cite{Fox:2002bu}. 
Here, we have suppressed the index $i=1,2,3$ for $g$, $m_D$, $m_A$, 
$b_A$ and $m_a^2$.

As we will see later, the relevant parameter region for us is where 
$\tan\beta \equiv \langle H_u \rangle/\langle H_d \rangle$ is not 
large, e.g. $\tan\beta \simlt 3$, so the only important Yukawa coupling 
is the top Yukawa coupling.  The soft supersymmetry breaking masses 
for the Higgs doublets are then given by
\begin{equation}
  m_{H_u}^2 \approx m_{\tilde{l}}^2 
    - \frac{3 y_t^2}{8\pi^2} (m_{\tilde{q}}^2+m_{\tilde{u}}^2) 
      \ln\biggl( \frac{m_{D,3}}{m_{\tilde{q}}} \biggr),
\qquad
  m_{H_d}^2 \approx m_{\tilde{l}}^2,
\label{eq:Higgs-soft}
\end{equation}
where we have used the fact that the mediation scale for the squark 
masses is of order the Dirac gluino mass $m_{D,3}$, and we have 
approximated $m_{\tilde{q}} \approx m_{\tilde{u}}$ inside the logarithm. 
A small soft mass squared for $S$, $m_S^2$, is also generated at one 
loop through $\lambda$, picking up $m_{H_u}^2$ and $m_{H_d}^2$. 
Equation~(\ref{eq:Higgs-soft}) explicitly demonstrates that the 
effective messenger scale for this theory is very low
\begin{equation}
  M_{\rm mess} \approx m_{D,3} 
    \approx \frac{\sqrt{3 \pi^2}}{g_3} m_{\tilde{q}},
\label{eq:eff-mess}
\end{equation}
so that larger squark masses can, in principle, be obtained for a given 
fine-tuning and Higgs boson mass, compared with gauge-mediation-type 
models such as the ones considered in~\cite{Chacko:2005ra,Nomura:2005qg}. 
This is because the squark squared masses are suppressed by a one-loop 
factor compared with the squared messenger scale in the present theory, 
while they are suppressed by a two-loop factor in gauge-mediation-type 
models.

Weak-scale values for the couplings $\lambda$ and $\kappa$ are subject 
to the constraint that they do not hit the Landau pole below the 
unification scale.  In our theory, the 321 gauge couplings are large 
at the ultraviolet due to the introduction of the $A$ fields and 
any additional fields, e.g., needed to recover coupling unification, 
which significantly weakens these constraints.  We find that $\lambda 
\simlt 0.8$ can be obtained for $\tan\beta \simgt 1.8$ for sufficiently 
large matter content, while the bound becomes somewhat stronger for 
smaller $\tan\beta$, e.g. $\lambda \simlt 0.7$ for $\tan\beta \simgt 
1.4$~\cite{Chacko:2005ra,Masip:1998jc}.  Note that, with the strong 
321 gauge couplings at the ultraviolet, $\tan\beta$ as small as 
$\sim 1.2$ is allowed because $y_t$ receives a strong asymptotically 
non-free contribution from a large $SU(3)_C$ coupling at the ultraviolet. 
The bound on $\kappa$ is given by $\kappa \simlt 0.2$ ($0.3$) for 
$\lambda \simeq 0.8$ ($0.7$).

\subsection{Suppression of the {\boldmath $D$}-term potential and 
a constraint from the {\boldmath $\rho$} parameter}
\label{subsec:1-D-pot}

The operators in Eq.~(\ref{eq:gaugino-mass}) give mixings between the 
auxiliary $D$ fields and the scalar components of $A$ (the third and 
fourth terms of Eq.~(\ref{eq:operators})).  As a consequence, the 
$SU(2)_L$ and $U(1)_Y$ $D$-term contributions to the Higgs quartic 
couplings are suppressed~\cite{Fox:2002bu}.  Denoting the suppression 
factors by $\epsilon$ ($\epsilon_2$ and $\epsilon_1$ for $SU(2)_L$ 
and $U(1)_Y$, respectively), they are given by
\begin{equation}
  \epsilon = \frac{m_A^2 - b_A + m_a^2}{4 g^2 m_D^2 
    + m_A^2 - b_A + m_a^2},
\label{eq:epsilon}
\end{equation}
where, again, we have suppressed the index $i=1,2$.  The $D$-term 
contributions to the Higgs potential are given by $\epsilon$ times 
the standard contributions.  

The suppression of the $D$-term potential can also be seen before 
integrating out the $A$ fields.  Focusing on the $T^3$ direction 
of $SU(2)_L$, the corresponding $D$-term potential is given, after 
integrating out the $D$ field, by
\begin{equation}
  V = \frac{g_2^2}{2} \biggl( \frac{1}{2} h_u^2 - \frac{1}{2} h_d^2 
    + 2 m_{D,2}\, \varphi_2 \biggr)^2,
\label{eq:D-pot}
\end{equation}
where $\varphi_2$ is the imaginary part of the $T^3$ component of 
the $SU(2)_L$ adjoint field, $a_2 = i \varphi_2/\sqrt{2} + \cdots$, 
and we have retained only the components for the Higgs doublets that 
obtain VEVs, $H_u = (0, h_u)^T$ and $H_d = (h_d, 0)^T$.  The potential 
of Eq.~(\ref{eq:D-pot}) forces $\varphi_2$ to have a VEV
\begin{equation}
  \langle \varphi_2 \rangle 
    = (1-\epsilon_2) \frac{\cos(2\beta)}{4 m_{D,2}} v^2
    \approx \frac{\cos(2\beta)}{4 m_{D,2}} v^2,
\label{eq:phi2-VEV}
\end{equation}
where $v^2 \equiv \langle h_u \rangle^2 + \langle h_d \rangle^2$ 
and $\epsilon_2$ is given by Eq.~(\ref{eq:epsilon}).  This mostly 
cancels the $D$-term potential.  (We find $V = (\epsilon_2 g_2^2/2) 
(\langle h_u\rangle^2/2 - \langle h_d \rangle^2/2)$ by substituting 
$\langle \varphi_2 \rangle$ back to $V$.) 

The size of any $SU(2)_L$ triplet VEV is subject to a stringent 
constraint from electroweak data (the $\rho$ parameter).  This 
gives the upper bound on the value of $\langle \varphi_2 \rangle$, 
and thus the lower bound on $m_{D,2}$.  Requiring $|\rho - 1| \approx 
\cos^2(2\beta)v^2/8 m_{D,2}^2 \simlt 0.002$~\cite{Eidelman:2004wy}, 
we find $m_{D,2} \simgt 1~{\rm TeV}$ for $\tan\beta \approx 2$. 
This bound is easily satisfied in the parameter region considered 
in the next subsection.

The imaginary part of the singlet field $a_1$ also receives a small 
VEV of order $v^2/m_{D,1}$, analogously to $\varphi_2$.  This VEV, 
however, does not affect phenomenology except that it is responsible 
for the suppression of the $U(1)_Y$ $D$-term potential.

\subsection{Electroweak symmetry breaking}
\label{subsec:1-ewsb}

We are now ready to discuss electroweak symmetry breaking.  Our Higgs 
potential is given by
\begin{eqnarray}
  V = V_F + V_D + V_{\rm soft},
\label{eq:Higgs-Pot}
\end{eqnarray}
where $V_F$, $V_D$ and $V_{\rm soft}$ are given by
\begin{eqnarray}
  V_F &=& |\lambda H_u H_d + M_S S + \kappa S^2|^2 
    + |\lambda S H_u + \mu H_u|^2 + |\lambda S H_d + \mu H_d|^2,
\label{eq:VF} \\
  V_D &=& \epsilon_2 \frac{g_2^2}{2} \sum_{a=1}^{3} 
    \biggl( H_u^\dagger \frac{\sigma^a}{2} H_u 
      + H_d^\dagger \frac{\sigma^a}{2} H_d \biggr)^2
    + \epsilon_1 \frac{3 g_1^2}{10} \biggl( \frac{1}{2} H_u^\dagger H_u 
      - \frac{1}{2} H_d^\dagger H_d \biggr)^2,
\label{eq:VD} \\
  V_{\rm soft} &=& m_{H_u}^2 |H_u|^2 + m_{H_d}^2 |H_d|^2 
    + \biggl( b_H H_u H_d + \frac{b_S}{2} S^2 + {\rm h.c.} \biggr).
\label{eq:Vsoft}
\end{eqnarray}
Here, $\epsilon_2$ and $\epsilon_1$ are given by Eq.~(\ref{eq:epsilon}), 
and $m_{H_u}^2$ and $m_{H_d}^2$ by Eq.~(\ref{eq:Higgs-soft}).  The 
holomorphic supersymmetry breaking masses $b_H$ and $b_S$, which we 
write as
\begin{eqnarray}
  b_H &=& \mu\, m_{3/2} - b_{H,0},
\label{eq:bH-1} \\
  b_S &=& M_S m_{3/2} - b_{S,0},
\label{eq:bS-1}
\end{eqnarray}
are given by Eq.~(\ref{eq:soft-mass-H-S}).  Other supersymmetry breaking 
parameters are also generated at higher loop orders.

The results of the potential minimization are given in 
Table~\ref{table:model1-points} for four sample points A, B, C and D, 
which lead to realistic phenomenology. 
\begin{table}
\begin{center}
\begin{tabular}{|c|c|c|c|c|}  \hline 
                             &    A    &    B    &    C    &    D    
\\ \hline
 $\lambda$                   &   $0.8$ &   $0.8$ &   $0.7$ &   $0.7$ \\
 $\kappa$                    &   $0.2$ &   $0.2$ &   $0.2$ &   $0.2$ \\
 $\mu$                       &   $177$ &   $226$ &   $186$ &   $236$ \\
 $M_S$                       &   $355$ &   $452$ &   $373$ &   $473$ \\
 $m_{3/2}$                   &   $183$ &   $234$ &   $225$ &   $285$ \\
 $[b_{H,0}]^{1/2}$           &  $-100$ &   $-93$ &    $41$ &  $-111$ \\
 $[b_{S,0}]^{1/2}$           &  $-133$ &  $-177$ &   $-86$ &  $-110$ 
\\ \hline
 $m_{D,1}$                   &  $5076$ &  $6935$ &  $5331$ &  $8112$ \\
 $m_{D,2}$                   &  $2201$ &  $2688$ &  $2312$ &  $3424$ \\
 $m_{D,3}$                   &  $2018$ &  $2571$ &  $2120$ &  $2690$ \\
 $m_{A,1}$                   &   $385$ &   $491$ &  $2242$ &  $2853$ \\
 $m_{A,2}$                   &   $269$ &   $343$ &  $2415$ &  $2527$ \\
 $m_{A,3}$                   &   $459$ &   $584$ &  $2312$ &  $2935$ \\
 $[b_{A,1}]^{1/2}$           &    $92$ &   $117$ &   $225$ &   $285$ \\
 $[b_{A,2}]^{1/2}$           &   $110$ &   $140$ &   $308$ &   $391$ \\
 $[b_{A,3}]^{1/2}$           &   $318$ &   $405$ &   $334$ &   $424$ 
\\ \hline
 $\tan\beta$                 &   $1.8$ &   $1.8$ &   $1.7$ &   $1.6$ \\
 $\mu_{\rm eff}$             &   $175$ &   $224$ &   $185$ &   $236$ \\
 $[(\mu B)_{\rm eff}]^{1/2}$ &   $204$ &   $246$ &   $199$ &   $282$ \\
 $[m_{H_u}^2]^{1/2}$         &  $-123$ &  $-159$ &  $-137$ &  $-117$ \\
 $[m_{H_d}^2]^{1/2}$         &   $170$ &   $213$ &   $153$ &   $244$ \\
 $M_{\rm Higgs}$             &   $134$ &   $138$ &   $128$ &   $135$ 
\\ \hline
 $\langle S \rangle$         &  $-2.8$ &  $-2.4$ &  $-2.1$ &  $-1.3$ \\
 $\epsilon_1$                & $0.0063$& $0.0054$&  $0.17$ &  $0.12$ \\
 $\epsilon_2$                & $0.0073$& $0.0080$&  $0.39$ &  $0.24$ 
\\ \hline
 $(m_{\tilde{q}}^2)^{1/2}$   &   $538$ &   $683$ &   $519$ &   $672$ \\
 $(m_{\tilde{u}}^2)^{1/2}$   &   $527$ &   $673$ &   $512$ &   $657$ \\
 $(m_{\tilde{d}}^2)^{1/2}$   &   $519$ &   $661$ &   $504$ &   $642$ \\
 $(m_{\tilde{l}}^2)^{1/2}$   &   $170$ &   $213$ &   $153$ &   $244$ \\
 $(m_{\tilde{e}}^2)^{1/2}$   &   $159$ &   $217$ &   $156$ &   $242$ 
\\ \hline
 $\tilde{\Delta}^{-1}$       &  $24\%$ &  $16\%$ &  $20\%$ &  $13\%$ 
\\ \hline
\end{tabular}
\end{center}
\caption{Values for the parameters of the model for four sample points, 
 A, B, C and D.  The resulting soft supersymmetry breaking masses for 
 squarks and sleptons as well as the quantities in the Higgs sector are 
 also listed.  Here, $[X]^{n} \equiv {\rm sgn}(X) \cdot |X|^{n}$, 
 and all masses are given in units of GeV.  The fine-tuning parameter 
 $\tilde{\Delta}^{-1}$ is defined in Ref.~\cite{Chacko:2005ra}.}
\label{table:model1-points}
\end{table}
The parameters $m_{D,i}$, $m_{A,i}$, and $b_{A,i}$ ($i=1,2,3$) are 
Dirac gaugino masses, supersymmetric masses for $A_i$, and holomorphic 
supersymmetry breaking masses for $A_i$, respectively, and defined 
below Eq.~(\ref{eq:Ai=0}), below Eq.~(\ref{eq:SUSY-mass-Ai}), and in 
Eq.~(\ref{eq:soft-mass-Ai}).  The square bracket in the table is defined 
as $[X]^{n} \equiv {\rm sgn}(X) \cdot |X|^{n}$, and all masses are 
given in units of GeV.  The effective $\mu$ and $\mu B$ parameters are 
defined by $\mu_{\rm eff} \equiv \mu + \lambda \langle S \rangle$ and 
$(\mu B)_{\rm eff} \equiv b_H + \lambda(M_S \langle S \rangle + \kappa 
\langle S \rangle^2)$, and $M_{\rm Higgs}$ is the lightest Higgs boson 
mass.  We also list the parameter $\tilde{\Delta}^{-1}$ defined in 
Ref.~\cite{Chacko:2005ra}, following~\cite{Anderson:1994dz}, as a measure 
of fine-tuning in our theory.  All the parameters in the Higgs potential 
are taken to be real.

Our procedure to obtain these numbers is as follows.  The input parameters 
of the analysis are $\lambda$, $\kappa$, $\mu$, $M_S$, $m_{3/2}$, $b_{H,0}$, 
$b_{S,0}$, $m_{D,i}$, $m_{A,i}$, and $b_{A,i}$.  Using these, we can derive 
$m_{H_u}^2$ and $m_{H_d}^2$, assuming some initial value for $\tan\beta$ 
(which will be determined in the end by iteration).  This determines the 
Higgs potential of Eq.~(\ref{eq:Higgs-Pot}).  We also add the one-loop 
contribution from top-stop loops to the Higgs quartic coupling in our 
analysis.  (A precise calculation of this contribution requires a knowledge 
of $\langle S \rangle$, determined by iteration, but the effect of 
$\langle S \rangle \neq 0$ is negligible.)  Corrections from higher loops 
are not so large for the values of top squark masses considered here, 
only giving an additional negative contribution to the lightest Higgs 
boson mass of order a few GeV.  By minimizing the potential, we obtain 
$\langle H_u \rangle$, $\langle H_d \rangle$ and $\langle S \rangle$. 
These VEVs do not in general satisfy $v_H^2 \equiv \langle H_u \rangle^2 
+ \langle H_d \rangle^2 = (174~{\rm GeV})^2 $, so we iterate the entire 
process again using the input parameters appropriately rescaled by powers 
of $(174~{\rm GeV}/v_H)$ according to their dimensions.  In this process 
we use the derived value of $\tan\beta$, $\tan\beta = \langle H_u 
\rangle/\langle H_d \rangle$ to determine $m_{H_u}^2$.  By iterating this 
several times, we obtain the final values for the parameters, which gives 
$\langle H_u \rangle^2 + \langle H_d \rangle^2 = (174~{\rm GeV})^2$. 
The convergence of the whole procedure is fairly quick.

As is seen in the table, the fine-tuning required in our theory is very 
mild and only of order $20\%$.  We find that the electroweak scale $v$ 
is mainly sensitive to the values of $\mu$, $m_{3/2}$, $m_{D,2}$, $m_{D,3}$, 
$g_2$, $g_3$ and $y_t$, and the fine-tuning parameter is determined by 
the sensitivity to $\mu$, $m_{3/2}$, $m_{D,3}$, $g_3$ and $y_t$ in most 
of the parameter region.  The reduction of the fine-tuning occurs mainly 
because the restoring force of the Higgs potential arises from the $F$-term 
potential, which is stronger than the one from the $SU(2) \times U(1)_Y$ 
$D$-term potential~\cite{Nomura:2005rk}.  A very small effective messenger 
scale of Eq.~(\ref{eq:eff-mess}) then allows squark masses as large as 
$(500\!\sim\!700)~{\rm GeV}$.  As can be seen from the table, electroweak 
symmetry breaking in our theory is caused by an interplay between the 
$\mu B$ term and the top-stop radiative correction to $m_{H_u}^2$ --- 
the diagonal entries in the Higgs boson mass-squared matrix, $\approx 
\{ |\mu_{\rm eff}|^2 + m_{H_u}^2, |\mu_{\rm eff}|^2 + m_{H_d}^2 \}$, are 
both positive, and one of the eigenvalues becomes negative because of 
a non-zero value of $(\mu B)_{\rm eff}$.

\subsection{Superparticle, Higgs boson and adjoint scalar spectrum}
\label{subsec:1-spectrum}

The masses for the superparticles, the Higgs bosons, and the 
adjoint scalars are calculated at tree level for the four sample 
points in Table~\ref{table:model1-points}, which are listed in 
Table~\ref{table:model1-spectra}. 
\begin{table}
\begin{center}
\begin{tabular}{|c|c|c|c|c|}  \hline 
                  &    A    &    B    &    C    &    D    
\\ \hline
 $\tilde{g}_1$    & $1982$  & $2526$  & $1427$  & $1812$  \\
 $\tilde{g}_2$    & $2441$  & $3110$  & $3740$  & $4747$  
\\ \hline
 $\chi^{\pm}_1$   &  $175$  &  $224$  &  $189$  &  $238$  \\
 $\chi^{\pm}_2$   & $1305$  & $1586$  &  $726$  & $1299$  \\
 $\chi^{\pm}_3$   & $1574$  & $1929$  & $3136$  & $3823$  
\\ \hline
 $\chi^0_1$       &  $169$  &  $219$  &  $186$  &  $236$  \\
 $\chi^0_2$       &  $207$  &  $250$  &  $208$  &  $255$  \\
 $\chi^0_3$       &  $392$  &  $482$  &  $400$  &  $494$  \\
 $\chi^0_4$       & $1305$  & $1586$  &  $726$  & $1299$  \\
 $\chi^0_5$       & $1574$  & $1929$  & $1599$  & $2605$  \\
 $\chi^0_6$       & $2175$  & $2987$  & $3136$  & $3823$  \\
 $\chi^0_7$       & $2560$  & $3478$  & $3840$  & $5458$  
\\ \hline
 $\tilde{u}_L$    &  $538$  &  $683$  &  $519$  &  $672$  \\
 $\tilde{u}_R$    &  $527$  &  $673$  &  $512$  &  $657$  \\
 $\tilde{d}_L$    &  $538$  &  $683$  &  $520$  &  $672$  \\
 $\tilde{d}_R$    &  $519$  &  $661$  &  $504$  &  $642$  \\
 $\tilde{e}_L$    &  $170$  &  $213$  &  $155$  &  $244$  \\
 $\tilde{e}_R$    &  $159$  &  $217$  &  $157$  &  $242$  \\
 $\tilde{\nu}_L$  &  $170$  &  $213$  &  $151$  &  $243$  
\\ \hline
 $\tilde{t}_1$    &  $521$  &  $654$  &  $505$  &  $634$  \\
 $\tilde{t}_2$    &  $561$  &  $696$  &  $545$  &  $686$  
\\ \hline
 $H^0_1$          &  $134$  &  $138$  &  $128$  &  $135$  \\
 $H^0_2$          &  $285$  &  $357$  &  $284$  &  $405$  \\
 $H^0_3$          &  $476$  &  $600$  &  $493$  &  $618$  \\
 $P^0_1$          &  $173$  &  $220$  &  $177$  &  $253$  \\
 $P^0_2$          &  $365$  &  $430$  &  $354$  &  $453$  \\
 $H^{\pm}$        &  $280$  &  $353$  &  $281$  &  $403$  
\\ \hline
 $a_{Y,1}$        &  $396$  &  $505$  & $2253$  & $2868$  \\
 $a_{Y,2}$        & $4733$  & $6464$  & $5434$  & $8057$  \\
 $a_{L,1}$        &  $291$  &  $370$  & $2435$  & $2557$  \\
 $a_{L,2}$        & $2872$  & $3509$  & $3844$  & $5104$  \\
 $a_{C,1}$        &  $558$  &  $711$  & $2336$  & $2965$  \\
 $a_{C,2}$        & $4412$  & $5622$  & $5156$  & $6544$  
\\ \hline
 $\tilde{G}$      &   $183$ &  $234$  &  $225$  &  $285$  
\\ \hline
\end{tabular}
\end{center}
\caption{The masses for the superparticles, Higgs bosons and 
 adjoint scalars for the four sample points A, B, C and D given in 
 Table~\ref{table:model1-points}.  All masses are given in units 
 of GeV.}
\label{table:model1-spectra}
\end{table}
The $\tilde{g}_{1,2}$, $\chi^{\pm}_{1,2,3}$, and $\chi^0_{1-7}$ 
represent the two gluinos, three charginos, and seven neutralinos, 
respectively, which come from the linear combinations of the original 
gauginos, $\lambda_i^\alpha$, and the fermionic components of $A_i$, $H_u$ 
and $H_d$.  The $\tilde{u}_{L,R}$, $\tilde{d}_{L,R}$, $\tilde{e}_{L,R}$, 
and $\tilde{\nu}_L$ represent the left- and right-handed up-type squarks, 
down-type squarks, charged sleptons, and the (left-handed) sneutrinos, 
respectively.  The masses for the top squarks, $\tilde{t}_{1,2}$, are 
listed separately because they split from the other squark masses 
appreciably.  The neutral scalar, pseudo-scalar, and charged Higgs bosons 
are labeled as $H^0_{1,2,3}$, $P^0_{1,2}$, and $H^{\pm}$, respectively, 
which arise from the scalar components of $S$, $H_u$ and $H_d$.  There 
are two adjoint scalar fields for each gauge group factor, which are 
denoted by $a_Y$, $a_L$, and $a_C$ for $U(1)_Y$, $SU(2)_L$, and $SU(3)_C$, 
respectively.  The gravitino is denoted by $\tilde{G}$.

In the first two points, A and B, the parameters are chosen such that 
$m_{A,i} \ll m_{D,i}$, so that the two gauginos for each gauge group 
factor are relatively close in masses: $\{ \tilde{g}_1, \tilde{g}_2 \}$ 
for $SU(3)_C$, $\{ \chi^{\pm}_2, \chi^{\pm}_3\}$ and $\{ \chi^0_4, 
\chi^0_5 \}$ for $SU(2)_L$, and $\{ \chi^0_6, \chi^0_7 \}$ for $U(1)_Y$. 
Because of small values for $m_{A,i}$, one of the two adjoint scalars 
for each gauge group factor, $a_{Y,1}$, $a_{L,1}$ and $a_{C,1}$, are 
relatively light (below a TeV).  For the other two points, C and D, the 
parameters are chosen such that $m_{A,i} \sim m_{D,i}$.  Thus the two 
gaugino masses are not necessarily close, and the adjoint scalars are 
all heavy, with masses above $2~{\rm TeV}$. 

An interesting feature of the present model is that the effects of the 
gauge $D$-terms are suppressed because of mixings between the auxiliary 
$D$ fields and the $a$ fields (see Eq.~(\ref{eq:operators})).  This also 
affects the spectrum of superparticles.  For points A and B, the $SU(2)_L$ 
and $U(1)_Y$ $D$-terms receive large suppressions, $\epsilon_1, \epsilon_2 
\ll 1$ (see Table~\ref{table:model1-points}).  As a consequence, the 
squarks and sleptons that are in the same $SU(2)_L$ multiplet are almost 
completely degenerate in mass.  (Mass splittings of order a few hundreds 
of MeV are generated from radiative corrections.)  For points C and D, the 
suppressions are not as strong as the case of points A and B, because of 
relatively large values of $m_{A,i}$, but the squarks and sleptons in the 
same $SU(2)_L$ multiplet are still quite degenerate. 

We find that the lightest supersymmetric particle (LSP) in our theory can 
either be the third generation right-handed slepton $\tilde{e}_R$, the third 
generation sneutrino $\tilde{\nu}_L$, the lightest neutralino $\chi^0_1$, 
or the gravitino $\tilde{G}$.  In either case, the mass of the LSP is 
naturally in the range $\approx (100\!\sim\!300)~{\rm GeV}$.  Because of 
rather small values for $\tan\beta$ and $M_{\rm mess}$, the masses of the 
third generation $\tilde{e}_R$ and $\tilde{\nu}_L$ are almost degenerate 
with those of the corresponding first-two generation particles.  For the 
case of the $\tilde{\nu}_L$ LSP, the left-handed selectrons $\tilde{e}_L$ 
will also be very close in mass, with the mass difference to $\tilde{\nu}_L$ 
only of order a few hundreds of MeV to a few GeV.  For the $\chi^0_1$ 
LSP, it is almost purely the Higgsino, so that the lightest chargino 
$\chi^{\pm}_1$ will be close in mass to $\chi^0_1$ with the mass 
difference of order a few GeV. 

The lightest Higgs boson in our theory cannot be heavier than about 
$140~{\rm GeV}$.  The mass of the charged Higgs boson is also bounded 
by $m_{H^{\pm}} \simlt 450~{\rm GeV}$, as the amount of fine-tuning is 
correlated with the charged Higgs boson mass~\cite{Nomura:2005rk}.  This 
may have some implications on the rate of the $b \rightarrow s \gamma$ 
process.  While the current theoretical estimates for this process still 
have some uncertainties~\cite{Borzumati:1998tg}, the positive sign for 
the effective $\mu$ parameter seems to be preferred over the other one, 
with which a partial cancellation between the charged Higgs boson and 
chargino contributions is possible.

\section{Models with Sequestered Gauge Mediation}
\label{sec:model-2}

In this section we present the second class of models.  We find that 
the fine-tuning is reduced to the level of $(10\!\sim\!20)\%$ in these 
models.

\subsection{Models}
\label{subsec:2-model}

Our basic idea here is the following.  We consider gauge mediation 
models, in which superparticle masses are generated by loops of messenger 
fields~\cite{Dine:1981gu,Dine:1994vc}.  In particular, we consider a model 
in which supersymmetric and supersymmetry breaking masses for the messenger 
fields do not possess any particular ``unified'' relations (this requires 
multiple singlets in the messenger sector)~\cite{Chacko:2005ra,Agashe:1997kn}. 
Now, suppose that all the MSSM fields together with the messenger fields 
are localized on a $(3+1)$-dimensional subspace in some higher dimensional 
spacetime, and that supersymmetry breaking occurs at some other subspace, 
which is transmitted to the messenger sector through some bulk interactions. 
In this case, we can push up the fundamental scale of supersymmetry breaking 
to an intermediate scale without affecting the gauge mediated spectrum 
for the MSSM superparticles.  On the other hand, supersymmetric masses 
of order the weak scale can be generated in the Higgs sector from the 
K\"ahler potential terms, as was the case in section~\ref{subsec:1-Higgs}. 
In fact, this structure was used in Ref.~\cite{Nomura:2000uw} to generate 
the $\mu$ term in gauge mediation models, where the coincidence of the 
scales for the $\mu$ term and for the superparticle masses was also 
naturally obtained.  Here we adopt the basic construction of this model 
to demonstrate our point. 

Let us consider $(4+1)$-dimensional spacetime with the extra dimension 
compactified on an $S^1/Z_2$ orbifold, $y: [0, 2\pi]$, where $y$ is the 
coordinate for the fifth dimension.  The size of the extra dimension 
we consider is small, only one or two orders of magnitude larger than 
the inverse of the fundamental scale, which is of order the Planck scale. 
We consider that supersymmetry is dynamically broken on the $y=\pi R$ 
brane at the scale $\Lambda$, and (some of) the fields participating 
in this dynamics are charged under a $U(1)_m$ gauge multiplet located 
in the bulk~\cite{Nomura:2000uw}.  Our messenger sector is localized 
on the $y=0$ brane.  Let us first consider only a single vector-like 
messenger ${\cal D}({\bf 3}^*, {\bf 1})_{1/3} + \bar{\cal D}({\bf 3}, 
{\bf 1})_{-1/3}$, where the numbers represent the 321 gauge quantum 
numbers.  The superpotential interactions in the messenger sector are 
then given by
\begin{equation}
  {\cal L} = \delta(y) \int\!d^2\theta\, 
    \biggl( k_E X E \bar{E} + \frac{f}{3} X^3 
    + k_{\cal D} X {\cal D} \bar{\cal D} \biggr) + {\rm h.c.},
\label{eq:messenger}
\end{equation}
where $X$, $E$ and $\bar{E}$ are singlets under 321, and the $U(1)_m$ 
charges for these fields are chosen as $E(+1)$, $\bar{E}(-1)$, $X(0)$, 
${\cal D}(0)$ and $\bar{\cal D}(0)$.  Supersymmetry breaking is mediated 
from the $y=\pi R$ brane to the $y=0$ brane through $U(1)_m$ gauge 
interactions, generating positive supersymmetry breaking squared masses 
of order $\approx (g_m^2/16\pi^2)^2\Lambda^2$ for $E$ and $\bar{E}$. 
Here, $g_m$ is the 4D $U(1)_m$ gauge coupling, which is naturally 
suppressed by the volume of the extra dimension.  These positive squared 
masses in turn generate a negative mass squared for $X$ through the 
coupling $k_E$, triggering the VEVs for the lowest and highest components 
of the $X$ chiral superfield: $\langle X \rangle \neq 0$ and $\langle 
F_X \rangle \neq 0$.  Note that, while the superpotential interactions 
of Eq.~(\ref{eq:messenger}) possess a $U(1)_R$ symmetry, it is explicitly 
broken by the trilinear scalar interactions arising from anomaly 
mediation~\cite{Randall:1998uk,Giudice:1998xp}, so that the dangerous 
Goldstone boson does not arise.  These VEVs then provide the supersymmetric 
and supersymmetry breaking masses for the messenger fields ${\cal D}$ and 
$\bar{\cal D}$: $M_{\cal D} = k_{\cal D} \langle X \rangle$ and $F_{\cal D} 
= k_{\cal D} \langle F_X \rangle$.  For $k_E \sim f \sim k_{\cal D} \sim 
O(1)$, the sizes of these masses are $M_{\cal D}^2 \sim F_{\cal D} \sim 
O(g_m^4 \Lambda^2/(16\pi^2)^3)$, so that we can naturally obtain $M_{\cal 
D}/\Lambda \sim \sqrt{F_{\cal D}}/\Lambda \sim (10^{-6}\!\sim\!10^{-5})$ 
for $g_m^2 \approx (10^{-3}\!\sim\!10^{-2})$, which is consistent with 
the volume suppression of $g_m$.  We thus take $\Lambda \approx 
10^{10}~{\rm GeV}$ and $M_{\cal D} \approx \sqrt{F_{\cal D}} \approx 
(10\!\sim\!100)~{\rm TeV}$ in our analysis.  As we have seen, this 
requires some coincidence of the scales but does not require fine-tuning. 

Our messenger sector also contains vector-like messenger fields other than 
${\cal D}$ and $\bar{\cal D}$.  In particular, to preserve the successful 
prediction for gauge coupling unification at the leading-log level, we 
introduce messenger fields in complete $SU(5)$ multiplets.  Specifically, 
we introduce $n_{\bf 5}$ pairs of $({\cal D}, {\cal L}) + (\bar{\cal D}, 
\bar{\cal L})$ and $n_{\bf 10}$ pairs of $({\cal Q}, {\cal U} ,{\cal E}) 
+ (\bar{\cal Q}, \bar{\cal U}, \bar{\cal E})$, where the 321 gauge 
quantum numbers of ${\cal Q}, {\cal U}, {\cal D}, {\cal L}$ and ${\cal E}$ 
are the same as the corresponding MSSM fields.  The numbers $n_{\bf 5}$ 
and $n_{\bf 10}$ are bounded by $n_{\bf 5} + 3n_{\bf 10} \simlt 5$ due 
to the Landau pole consideration for the 321 gauge couplings.  We 
consider that each component of the messenger fields has independent 
supersymmetric and supersymmetry breaking masses $M$ and $F$: for example, 
we treat $M_{\cal D}$, $F_{\cal D}$, $M_{\cal L}$ and $F_{\cal L}$ to be 
all independent for $(n_{\bf 5}, n_{\bf 10})=(1,0)$.  There are a number 
of ways to achieve this.  The easiest way is to introduce $E$, $\bar{E}$ 
and $X$ fields as well as the interactions of Eq.~(\ref{eq:messenger}) for 
each messenger field.  Such a structure can naturally arise if we introduce 
a discrete $Z_3$ symmetry for each component of the messenger fields.%
\footnote{These structures are consistent with gauge unification if the 
unified symmetry is realized in higher dimensions~\cite{Kawamura:2000ev}.}
In any event, with these most general $M$'s and $F$'s, the gaugino masses, 
$M_a$, and the sfermion masses, $m_{\tilde{f}}$, at the messenger scale 
are written as
\begin{equation}
  M_i = \frac{g_i^2}{16\pi^2} \Lambda_{G,i},
\label{eq:gaugino-gmsb}
\end{equation}
and
\begin{equation}
  m_{\tilde{f}}^2 = 2\!\! \sum_{i=1,2,3} 
    \biggl(\frac{g_i^2}{16\pi^2}\biggr)^2 C_i^{\tilde{f}} \Lambda_{S,i}^2, 
\label{eq:sfermion-gmsb}
\end{equation}
where $i=1,2,3$ represents $U(1)_Y$, $SU(2)_L$ and $SU(3)_C$, and 
$C_i^{\tilde{f}}$ are the group theoretical factors.  The parameters 
$\Lambda_{G,i}$ and $\Lambda_{S,i}$ are of order $(10\!\sim\!100)~{\rm TeV}$, 
which can be explicitly calculated in terms of the $M$'s and $F$'s once the 
field content for the messengers is specified. 

The Higgs sector of the present model is essentially the same as the 
one in the previous model (see section~\ref{subsec:1-Higgs}).  The field 
content is given by $S({\bf 1}, {\bf 1})_0$, $H_u({\bf 1}, {\bf 2})_{1/2}$ 
and $H_d({\bf 1}, {\bf 2})_{-1/2}$.  Imposing the discrete $Z_{4,R}$ 
symmetry of Table~\ref{table:Z4R}, the effective superpotential arises 
both from $W_0 = \lambda S H_u H_d + (\kappa/3) S^3$ and $K = \lambda_H 
H_u H_d + (\lambda_S/2) S^2 + {\rm h.c.}$ as 
\begin{equation}
  W_H = \lambda S H_u H_d + \mu H_u H_d 
    + \frac{M_S}{2} S^2 + \frac{\kappa}{3} S^3,
\label{eq:W_H-2}
\end{equation}
where $\mu$ and $M_S$ are parameters of order the weak scale generated via 
the mechanism of~\cite{Giudice:1988yz}.  Note that this mechanism works even 
if the supersymmetry breaking sector (at the $y=\pi R$ brane) and the Higgs 
sector (at the $y=0$ brane) are geometrically separated~\cite{Nomura:2000uw}. 
The holomorphic supersymmetry breaking terms 
\begin{equation}
  {\cal L}_{H,{\rm soft}} \equiv 
    - b_H H_u H_d - \frac{b_S}{2} S^2 + {\rm h.c.}
\label{eq:soft-mass-H-S-2}
\end{equation}
are also generated from the K\"ahler potential terms as
\begin{eqnarray}
  b_H &=& \mu\, m_{3/2},
\label{eq:bH-2} \\
  b_S &=& M_S m_{3/2},
\label{eq:bS-2}
\end{eqnarray}
where $m_{3/2} \approx \Lambda^2/M_{\rm Pl}$ is the gravitino mass of order 
the weak scale.  The Yukawa couplings for the quark and lepton superfields 
are given by $W = y_u Q U H_u + y_d Q D H_d + y_e L E H_d$. 

We here note that a theory having essentially the same properties 
can also be formulated in warped spacetime of~\cite{Randall:1999ee}. 
We can simply make our $S^1/Z_2$ extra dimension warped, with the 
scales on the ultraviolet and infrared branes set to be around the 4D 
Planck scale and the intermediate scale, respectively.  The MSSM fields, 
the singlet field $S$, and fields in the messenger sector are all 
localized on the ultraviolet brane, while the $U(1)_m$ gauge multiplet 
propagates in the bulk.  Supersymmetry breaking occurs on the infrared 
brane, which is transmitted to the $E$ and $\bar{E}$ fields on the 
ultraviolet brane through bulk $U(1)_m$ gauge interactions, as in 
the models of~\cite{Gherghetta:2000qt,Goldberger:2002pc}.  This 
theory allows a purely 4D interpretation through the AdS/CFT 
correspondence~\cite{Maldacena:1997re,Arkani-Hamed:2000ds}, in which 
the separation of supersymmetry breaking and the other fields occurs 
through conformal sequestering effects~\cite{Luty:2001jh}.

\subsection{Electroweak symmetry breaking and particle spectrum}
\label{subsec:2-ewsb}

Now we study electroweak symmetry breaking in our model.  The Higgs 
potential is given by Eq.~(\ref{eq:Higgs-Pot}) with Eqs.~(\ref{eq:VF}~%
--~\ref{eq:Vsoft}), but with both $\epsilon_1$ and $\epsilon_2$ set 
to $1$ in Eq.~(\ref{eq:VD}).  The holomorphic supersymmetry breaking 
masses, $b_H$ and $b_S$, are given by Eqs.~(\ref{eq:bH-2},~\ref{eq:bS-2}) 
rather than Eqs.~(\ref{eq:bH-1},~\ref{eq:bS-1}).  For smaller number 
of messenger fields, the 321 gauge couplings are not very strong at 
the unification scale, so that the value of $\lambda$ should be somewhat 
smaller than $0.8$ to avoid the Landau pole. 

The results of the potential minimization are given in 
Table~\ref{table:model2-points} for three sample points A, B and C, 
which lead to realistic phenomenology. 
\begin{table}
\begin{center}
\begin{tabular}{|c|c|c|c|}  \hline 
                             &    A    &    B    &    C    
\\ \hline
 $\lambda$                   &  $0.65$ &  $0.75$ &  $0.75$ \\
 $\kappa$                    &   $0.2$ &   $0.2$ &   $0.2$ \\
 $\mu$                       &   $202$ &   $135$ &   $175$ \\
 $M_S$                       &   $598$ &   $580$ &   $616$ \\
 $m_{3/2}$                   &   $268$ &   $362$ &   $127$ 
\\ \hline
 $n_{\bf 5}$                 &     $1$ &     $4$ &     $1$ \\
 $n_{\bf 10}$                &     $0$ &     $0$ &     $1$ \\
 $M_{\rm mess}$              &    $50$ &    $50$ &    $50$ \\
 $\Lambda_{G,1}$             &    $53$ &   $100$ &   $150$ \\
 $\Lambda_{G,2}$             &    $68$ &   $140$ &    $82$ \\
 $\Lambda_{G,3}$             &    $30$ &    $39$ &    $48$ \\
 $\Lambda_{S,1}$             &    $56$ &    $57$ &    $98$ \\
 $\Lambda_{S,2}$             &    $68$ &    $72$ &    $49$ \\
 $\Lambda_{S,3}$             &    $30$ &    $19$ &    $24$ 
\\ \hline
 $\tan\beta$                 &   $1.9$ &   $1.8$ &   $3.6$ \\
 $\mu_{\rm eff}$             &   $204$ &   $139$ &   $175$ \\
 $[(\mu B)_{\rm eff}]^{1/2}$ &   $234$ &   $224$ &   $149$ \\
 $[m_{H_u}^2]^{1/2}$         &  $-143$ &    $17$ &  $-177$ \\
 $[m_{H_d}^2]^{1/2}$         &   $238$ &   $250$ &   $187$ \\
 $\langle S \rangle$         &   $1.6$ &   $4.4$ &  $-0.7$ \\
 $M_{\rm Higgs}$             &   $123$ &   $129$ &   $120$ 
\\ \hline
 $M_1$                       &    $74$ &   $142$ &   $213$ \\
 $M_2$                       &   $185$ &   $391$ &   $224$ \\
 $M_3$                       &   $258$ &   $333$ &   $411$ \\
 $(m_{\tilde{q}}^2)^{1/2}$   &   $450$ &   $406$ &   $434$ \\
 $(m_{\tilde{u}}^2)^{1/2}$   &   $391$ &   $335$ &   $416$ \\
 $(m_{\tilde{d}}^2)^{1/2}$   &   $387$ &   $331$ &   $406$ \\
 $(m_{\tilde{l}}^2)^{1/2}$   &   $238$ &   $250$ &   $187$ \\
 $(m_{\tilde{e}}^2)^{1/2}$   &    $94$ &    $95$ &   $165$ 
\\ \hline
 $\tilde{\Delta}^{-1}$       &  $13\%$ &  $19\%$ &  $12\%$ 
\\ \hline
\end{tabular}
\end{center}
\caption{Values for the parameters of the model for three sample points, 
 A, B and C.  The resulting soft supersymmetry breaking masses for the 
 gauginos, squarks and sleptons, as well as the quantities in the Higgs 
 sector, are also listed.  Here, $[X]^{n} \equiv {\rm sgn}(X) \cdot 
 |X|^{n}$.  All masses are given in units of GeV except for $M_{\rm mess}$, 
 $\Lambda_{G,i}$ and $\Lambda_{S,i}$ ($i=1,2,3$), which are given in units 
 of TeV.  The fine-tuning parameter $\tilde{\Delta}^{-1}$ is defined in 
 Ref.~\cite{Chacko:2005ra}.}
\label{table:model2-points}
\end{table}
The square bracket in the table is defined as $[X]^{n} \equiv 
{\rm sgn}(X) \cdot |X|^{n}$, and all masses are given in units of GeV 
except for $M_{\rm mess}$, $\Lambda_{G,i}$ and $\Lambda_{S,i}$, which are 
given in units of TeV.  The parameters $\Lambda_{G,i}$ and $\Lambda_{S,i}$ 
are defined in Eqs.~(\ref{eq:gaugino-gmsb},~\ref{eq:sfermion-gmsb}).  The 
quantity $M_{\rm mess}$ represents the scale at which the gaugino and 
sfermion masses of Eqs.~(\ref{eq:gaugino-gmsb},~\ref{eq:sfermion-gmsb}) are 
given, which we take as a single scale of order $\Lambda$'s for simplicity. 
The sensitivity of physical quantities to this parameter is rather weak. 
The effective $\mu$ and $\mu B$ parameters are defined by $\mu_{\rm eff} 
\equiv \mu + \lambda \langle S \rangle$ and $(\mu B)_{\rm eff} \equiv 
b_H + \lambda(M_S \langle S \rangle + \kappa \langle S \rangle^2)$, 
and $M_{\rm Higgs}$ is the lightest Higgs boson mass.  We also 
list the fine-tuning parameter $\tilde{\Delta}^{-1}$ defined 
in~\cite{Chacko:2005ra}.  All the parameters in the Higgs potential 
are taken to be real.  The procedure to obtain these numbers is 
analogous to that in section~\ref{subsec:1-ewsb}. 

As is seen in the table, we find that the fine-tuning in this theory 
is at the level of $(10\!\sim\!20)\%$.  A difference from the previous 
model is that the logarithm $\ln(M_{\rm mess}/m_{\tilde{t}})$ appearing 
in the top-stop correction to the Higgs mass-squared parameter is now 
not as small as the previous one.  ($M_{\rm mess}$ is several tens of 
TeV in the present model while it is a few TeV in the previous model.) 
We find that for most of the parameter region the total mass-squared 
parameters for the up-type and down-type Higgs doublets are both positive, 
and electroweak symmetry breaking is triggered by a nonzero value of the 
effective $\mu B$ term, although the point~C has a negative squared 
mass for the up-type Higgs field, $\mu_{\rm eff}^2 + m_{H_u}^2 < 0$. 
While the reduction of fine-tuning in the present model is not as 
large as the previous one, the situation is still much better than 
in conventional models of supersymmetry breaking, which typically 
require fine-tuning of order a few percent or even worse.

We have listed the masses for the superparticles and the Higgs bosons 
in Table~\ref{table:model2-spectra} for the three sample points of 
Table~\ref{table:model2-points}. 
\begin{table}
\begin{center}
\begin{tabular}{|c|c|c|c|}  \hline 
                  &    A    &    B    &    C    
\\ \hline
 $\tilde{g}$      &  $306$  &  $376$  &  $450$  
\\ \hline
 $\chi^{\pm}_1$   &  $183$  &  $148$  &  $165$  \\
 $\chi^{\pm}_2$   &  $235$  &  $404$  &  $257$  
\\ \hline
 $\chi^0_1$       &   $84$  &  $158$  &  $187$  \\
 $\chi^0_2$       &  $204$  &  $158$  &  $207$  \\
 $\chi^0_3$       &  $204$  &  $159$  &  $207$  \\
 $\chi^0_4$       &  $216$  &  $365$  &  $216$  \\
 $\chi^0_5$       &  $616$  &  $606$  &  $641$  
\\ \hline
 $\tilde{u}_L$    &  $448$  &  $404$  &  $431$  \\
 $\tilde{u}_R$    &  $390$  &  $334$  &  $415$  \\
 $\tilde{d}_L$    &  $452$  &  $408$  &  $438$  \\
 $\tilde{d}_R$    &  $388$  &  $331$  &  $407$  \\
 $\tilde{e}_L$    &  $240$  &  $253$  &  $192$  \\
 $\tilde{e}_R$    &  $100$  &  $100$  &  $170$  \\
 $\tilde{\nu}_L$  &  $232$  &  $246$  &  $178$  
\\ \hline
 $\tilde{t}_1$    &  $358$  &  $310$  &  $384$  \\
 $\tilde{t}_2$    &  $451$  &  $415$  &  $448$  
\\ \hline
 $H^0_1$          &  $123$  &  $129$  &  $120$  \\
 $H^0_2$          &  $358$  &  $335$  &  $265$  \\
 $H^0_3$          &  $730$  &  $754$  &  $695$  \\
 $P^0_1$          &  $309$  &  $239$  &  $243$  \\
 $P^0_2$          &  $503$  &  $460$  &  $588$  \\
 $H^{\pm}$        &  $358$  &  $332$  &  $271$  
\\ \hline
 $\tilde{G}$      &  $268$  &  $362$  &  $127$  
\\ \hline
\end{tabular}
\end{center}
\caption{The masses for the superparticles and the Higgs bosons for the 
 three sample points A, B and C given in Table~\ref{table:model2-points}. 
 All masses are given in units of GeV.}
\label{table:model2-spectra}
\end{table}
Here, we have included one-loop threshold corrections to obtain these 
masses, as they are relevant for colored particles especially if the 
masses are close to the experimental bounds.  The meaning of the symbols 
is the same as that in Table~\ref{table:model2-spectra}: $\tilde{g}$, 
$\chi^{\pm}_{1,2}$ and $\chi^0_{1-5}$ denote the gluino, charginos 
and neutralinos, respectively, $\tilde{u}_{L,R}$, $\tilde{d}_{L,R}$, 
$\tilde{e}_{L,R}$ and $\tilde{\nu}_L$ the squarks and sleptons, and 
$H^0_{1,2,3}$, $P^0_{1,2}$ and $H^{\pm}$ the neutral-scalar, pseudo-scalar, 
and charged Higgs bosons, respectively.  The top squarks, $\tilde{t}_{1,2}$, 
are listed separately, and $\tilde{G}$ is the gravitino.  As is seen in 
the table, the theory accommodates various possibilities for the LSP. 
We find that the LSP in this theory can be either the lightest neutralino 
$\chi^0_1$, the lightest chargino $\chi^{\pm}_1$, the third generation 
right-handed slepton $\tilde{e}_R$, the third generation sneutrino 
$\tilde{\nu}_L$, or the gravitino $\tilde{G}$.  We note that the theory 
allows the relative signs between $\Lambda_{G,i}$'s to be negative, 
although we did not adopt such a case in our sample points A, B and C. 
In the case that $\chi^0_1$ is the LSP, its thermal relics may provide 
the dark matter of the universe~\cite{Nomura:2001ub}.  In general, the 
neutral LSP's, $\chi^0_1$ and $\tilde{G}$, can be the dark matter if 
they are nonthermally produced.  The cases with charged LSP's require 
``non-standard'' cosmology or small $R$-parity violation.

\section{Conclusions}
\label{sec:concl}

We have constructed two classes of realistic supersymmetric models 
in which no significant fine-tuning is required to reproduce the 
correct scale for electroweak symmetry breaking.  Both classes of models 
accomplish this by incorporating (i) a means to increase the effective 
Higgs quartic coupling beyond what is available in the MSSM, (ii) a low 
scale for the generation of squark and slepton masses, and (iii) a 
degree of independent adjustability for the size of the squark masses, 
such that they can be set not very far from their experimental bound 
of $\approx 300~{\rm GeV}$.  The first feature allows us to evade 
the LEP~II bound of $M_{\rm Higgs} \simgt 114~{\rm GeV}$ without 
relying on large radiative corrections from top-stop loops, and thus 
to have small top squark masses.  The second and third features then 
facilitate a softening of the top-stop loop correction to the up-type 
Higgs mass-squared parameter to a level commensurable with the Higgs 
field VEV $v \simeq 174~{\rm GeV}$, thereby eliminating the need for 
delicate cancellations.  Specifically, the low mass-generation scale 
leads to a modest logarithm, $\simeq (2\!\sim\!5)$, in the correction, 
and the free adjustability of the squark masses moderates the overall 
mass scale of the correction.

In the first class of models, the additional Higgs quartic 
contribution is provided by $F$-exchange of a singlet chiral superfield.  
A fundamental intermediate-scale supersymmetry breaking is communicated 
to the MSSM fields primarily via two means.  First, a new $U(1)$ gauge 
superfield acquires a $D$-term VEV at the intermediate scale and marries 
the 321 gauginos with adjoint chiral fermions via nonrenormalizable 
interactions.  This leads to a consequence that the radiative generation 
of the sfermion masses occurs at a scale only a factor of $O(4\pi)$ 
higher than the electroweak scale.  This accomplishes the low scale 
of sfermion mass generation.  The adjustability of squark masses 
is accomplished simply by making the nonrenormalizable couplings 
responsible for the gaugino masses independent for $SU(3)_C$, $SU(2)_L$, 
and $U(1)_Y$.  The second means for transmitting supersymmetry breaking 
is via supergravity effects.  These naturally generate weak-scale 
supersymmetric and supersymmetry breaking masses for the Higgs and 
singlet fields, producing an appropriate Higgs sector superpotential. 
Working together, these characteristic features of the models 
allow for the coexistence of the electroweak VEV with a variety 
of superparticle spectra, with a very mild tuning of about $20\%$.

The second class of models also relies on $F$-exchange of a singlet 
to obtain the additional Higgs quartic coupling.  However, communication 
of supersymmetry breaking to the MSSM sector more closely follows 
traditional gauge mediation.  This leads to sfermion mass-generation 
scales $O(16\pi^2)$ higher than the electroweak scale, which are larger 
than the mass-generation scales for the first class of models, but 
still provides only modest logarithmic enhancement in the Higgs mass 
correction.  Supersymmetry is fundamentally broken at an intermediate
scale, at a location physically separated from the MSSM fields in 
an extra dimension.  The breaking is then communicated between the 
two locations by a bulk $U(1)$, which ultimately gives supersymmetric 
and supersymmetry breaking masses of $O(10\!\sim\!100~{\rm TeV})$ 
to the messenger fields of gauge mediation.  We assume no specific 
relation between any of the mass parameters for the messengers, 
ensuring the adjustability of the squark masses.  Weak-scale values 
for supersymmetric and supersymmetry breaking masses for the Higgs and 
singlet fields are again generated from supergravity effects.  Within 
this class of models the reduction of tunings to $(10\!\sim\!20)\%$ 
can be obtained, again with a variety of configurations for the 
superparticle spectrum.

In searching out parameter points of reduced tuning in the above 
models, it quickly becomes apparent that the weakest tunings and 
largest Higgs boson masses are generally obtained when the Higgs 
potential is made stable along both the $H_u$ and $H_d$ axes, and 
destabilized in an intermediate direction via the $\mu B$ term.  The 
reason is simply that the Higgs quartic coupling due to $F$-exchange 
is strongest when $\tan\beta \sim 1$.  Thus, to most successfully 
utilize the new quartic coupling, smaller values for $\tan\beta$ are 
preferred, and electroweak symmetry breaking must be dominantly caused 
by the $\mu B$ term.  This clearly demonstrates the framework introduced 
in~\cite{Nomura:2005rk}.  It is very encouraging that this very generic 
idea has been successfully demonstrated in two distinct classes of 
realistic models.

\section*{Acknowledgments}

This work was supported in part by the Director, Office of Science, Office 
of High Energy and Nuclear Physics, of the US Department of Energy under 
Contract DE-AC02-05CH11231.  The work of Y.N. was also supported by the 
National Science Foundation under grant PHY-0403380, by a DOE Outstanding 
Junior Investigator award, and by an Alfred P. Sloan Research Fellowship.

\end{document}